\documentclass[11pt]{article}
\usepackage{blindtext}
\usepackage[a4paper, total={6in, 8in}]{geometry}

\usepackage{diagbox}
\usepackage{graphicx}%
\usepackage{multirow}%
\usepackage{amsmath,amssymb,amsfonts}%
\usepackage{amsthm}%
\usepackage{mathrsfs}%
\usepackage[title]{appendix}%
\usepackage{xcolor}%
\usepackage{textcomp}%
\usepackage{manyfoot}%
\usepackage{booktabs}%
\usepackage{algorithm}%
\usepackage{algorithmicx}%
\usepackage{algpseudocode}%
\usepackage{listings}%
\usepackage{multirow}
\usepackage{authblk} 
\usepackage{caption}
\usepackage{float} 

\usepackage{hyperref}
\begin{document}

\title{Active Learning for Predicting the Enthalpy of Mixing in Binary Liquids Based on Ab Initio Molecular Dynamics}


\author[1,a]{Quentin Bizot\thanks{quentin.bizot@rub.de}}
\author[2,3]{Ryo Tamura\thanks{tamura.Ryo@nims.go.jp}}
\author[1]{Guillaume Deffrennes\thanks{guillaume.deffrennes@cnrs.fr}}

\affil[1]{Univ. Grenoble Alpes, CNRS, Grenoble INP, SIMaP, F-38000, Grenoble, France}
\affil[2]{Center for Basic Research on Materials, National Institute for Materials Science, 1-1 Namiki, Tsukuba, Ibaraki, 305-0044, Japan}
\affil[3]{Graduate School of Frontier Sciences, The University of Tokyo, 5-1-5 Kashiwa-no-ha, Kashiwa, Chiba, 277-8561, Japan}
\affil[a]{\textit{Present address of Q. Bizot:} Interdisciplinary Centre for Advanced Materials Simulations (ICAMS), Ruhr Universität Bochum, Bochum, Germany}

\maketitle

\abstract{
The enthalpy of mixing in the liquid phase is an important property for predicting phase formation in alloys. It can be estimated in a large compositional space from pairwise interactions between elements, for which machine learning has recently provided the most accurate predictions. Further improvements requires acquiring high-quality data in liquids where models are poorly constrained.
In this study, we propose an active learning approach to identify in which liquids additional data are most needed to improve an initial dataset that covers over 400 binary liquids. We identify a critical need for new data on liquids containing refractory elements, which we address by performing ab initio molecular dynamics simulations for 29 equimolar alloys of Ir, Os, Re and W. This enables more accurate predictions of the enthalpy of mixing
, and we discuss the trends obtained for refractory elements of period 6. We use clustering analysis to interpret the results of active learning and to explore how our features can be linked to Miedema's semi-empirical theory.
}

\vspace{1em}


\pagebreak

\section{Introduction}\label{sec1}



The enthalpy of mixing in the liquid phase is a property of fundamental importance for materials development. Its knowledge is essential for predicting the melting temperature range of alloys using Calphad \cite{kattner_calphad_2016} or machine learning \cite{deffrennes_framework_2023} models, which is important for the design of a variety of alloys, such as solders \cite{kattner_calculation_2001}, cast alloys \cite{sanchez_high-throughput_2025}, or phase change materials for thermal energy storage \cite{rawson_selection_2020}. This property also reflects the affinity between elements and serves as input in a variety of empirical and machine learning models, such as those used for the design of metallic glasses \cite{takeuchi_classification_2005}, the synthesis of high-entropy alloy nanoparticles \cite{cao_liquid_2023}, or the prediction of phase formation and mechanical properties in superalloys \cite{han_facilitated_2024} or high entropy alloys \cite{zhou_machine_2019, peivaste_data-driven_2023, rickman_materials_2019}.

In metallic liquids, the determination of the enthalpy of mixing in binary systems is the primary task, because reasonable extrapolations can be obtained in multicomponent liquids from the pairwise interactions \cite{deffrennes_data-driven_2024}. However, based on an extensive literature review \cite{deffrennes_data-driven_2024}, we estimate that measurements are only available in one in five binary systems, and in about one in two binary systems for the most studied elements such as copper or tin. Therefore, predicting the enthalpy of mixing in binary liquids remains highly important. To this day, Miedema's semi-empirical model \cite{miedema_cohesion_1980} has been regarded as the most reliable for this task. In a recent study, we developed a machine learning model that gives more accurate estimates at the cost of a lower interpretability \cite{deffrennes_data-driven_2024}. This model was trained on data collected in 375 binary systems from Calphad assessments, at compositions where direct or indirect measurements are available. While this represents an unprecedented amount of critically evaluated data, it covers only 17\% of the 2145 binary systems formed from the 66 elements in the dataset where the enthalpy of mixing is often only known in a limited range of composition. Besides, some elements, such as the low melting point metals, are well represented in the dataset, while others, like refractory or volatile elements, are underrepresented due to the scarcity and difficulty of obtaining experimental data. The performance of the machine learning model proposed in Ref. \cite{deffrennes_data-driven_2024}, as well as those being developed in subsequent studies \cite{huang_prediction_2025, vincely_amending_2025}, is therefore limited by the comprehensiveness and diversity of the training data. Reliable enthalpy of mixing data can be obtained by experimental methods such as calorimetry \cite{sommer_thermodynamics_2007}, or as we will demonstrate by ab initio molecular dynamics (AIMD), but in both cases this requires significant time and expertise. This raises a question of general importance: in which systems and compositions is new data on the enthalpy of mixing in the liquid phase needed the most for improving our understanding of this property? The answer is not simple, since previous attempts to predict liquidus \cite{deffrennes_framework_2023} and enthalpies of mixing \cite{deffrennes_data-driven_2024} have shown that accurate predictions can be achieved from limited data for some elements, such as rare earths, while for others, like aluminum, they remain challenging even with extensive data.

In this study, we propose an active learning strategy for selecting binary liquids for AIMD simulations, with the aim of efficiently improving the accuracy of machine learning predictions of the enthalpy of mixing. The layout of this paper is as follows. In Section \ref{Results}, we first presents our active learning strategy. Second, we report AIMD simulations results obtained for liquids of refractory metals. Third, we demonstrate the resulting improvements of our machine learning model. Fourth, we discuss the trends in the enthalpy of mixing for refractory elements. Last, we discuss a potential link between our features and Miedema's empirical parameters. Section \ref{Discussion} summarizes the key findings of this work, discusses the limitations of our model, and outlines future research directions. Finally, Section \ref{Methods} describes algorithms implementation and machine learning methods, as well as the protocol used for AIMD simulations.

\clearpage
\section{Results}
\label{Results}

\subsection{Identification of knowledge gaps using active learning and clustering}\label{ActiveLearningresults}

Using our initial dataset, we evaluated different strategies for identifying in which binary liquids the acquisition of new enthalpy of mixing data would be most informative: data acquisition in equimolar liquids selected at random, which serves as a benchmark, Gaussian process–based active learning, and regression tree–based active learning as proposed in Ref. \cite{jose_regression_2024}. Gaussian process–based active learning using 10 features selected by recursive feature elimination gave the best results, as detailed in Supplementary Notes 1 and 2. The 10 features are obtained from the composition and the properties of the pure elements: the composition-weighted average deviation between their group (1), heat capacity in the liquid at the melting point (2), heat capacity of melting (3), first ionization energy (4), enthalpy of melting (5), entropy of melting (6), and density (7), and the composition-weighted average of their group (8), heat capacity in the liquid at the melting point (9), and entropy of melting (10).

\begin{figure}[H]
\centering
\includegraphics[width=0.9\textwidth]{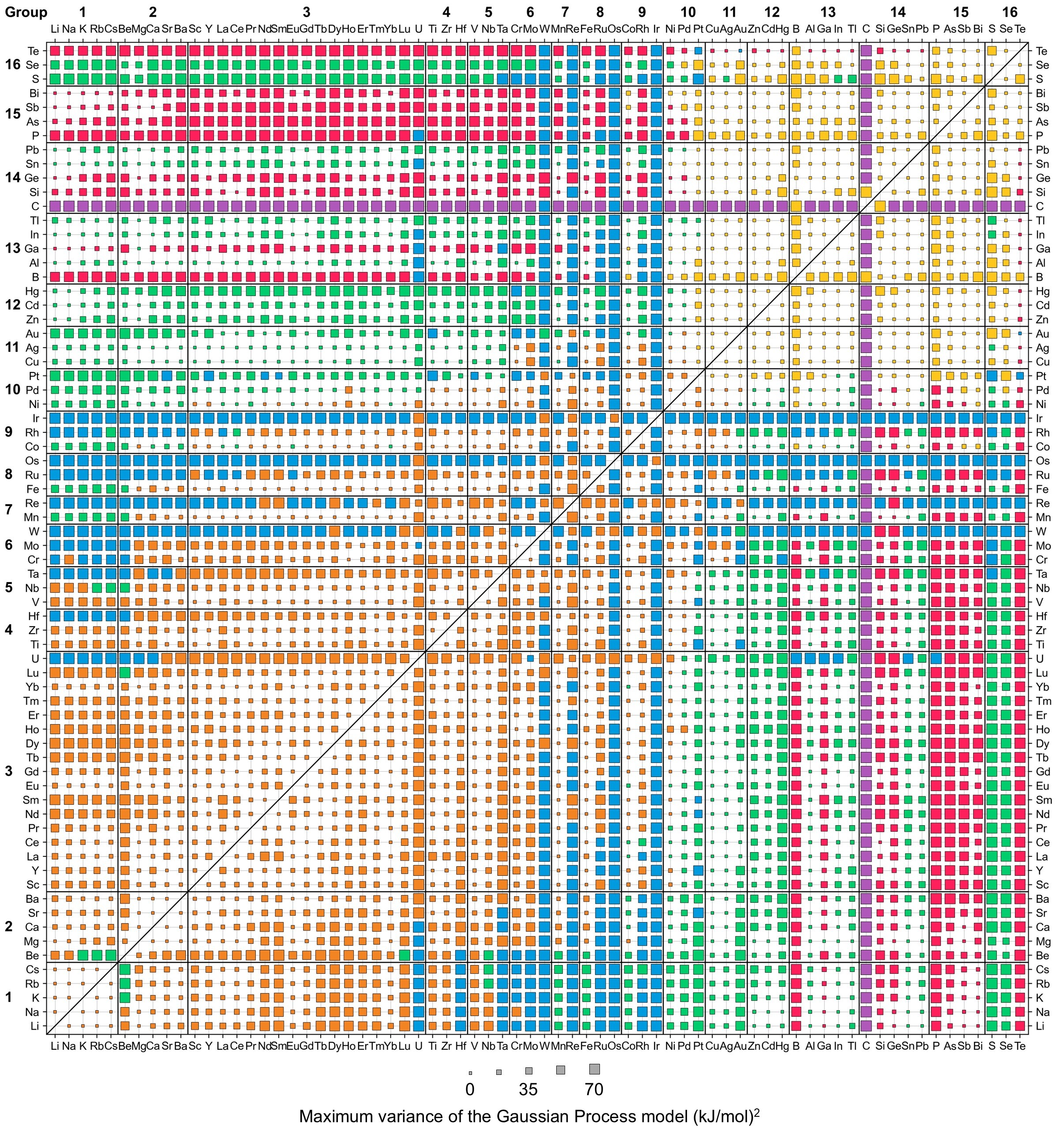}
\caption{Map of binary liquids with elements sorted from group 1 to 16. The size of the square represents the maximum variance in the enthalpy of mixing prediction generated by our Gaussian process model in the system, and its color the cluster to which the system belongs.}\label{fcov}
\end{figure}

Fig.\ref{fcov} shows the maximum variance of the Gaussian process, trained on the initial dataset, across 2415 binaries systems generated by 70 elements. Larger squares indicate a high uncertainty, and thus, that new data are needed. To better understand how model uncertainty varies between systems, we coupled our approach with clustering. We used K-means clustering to classify the binary systems into an optimal number of 6 groups, represented by the different colors in Fig.\ref{fcov}, based on the position of the equimolar liquid in the 10-dimensional feature space of the Gaussian process model. As a first observation, binary liquids containing carbon form a cluster of their own (purple). This can be understood by the fact that carbon has an enthalpy of melting at least twice as high as any of the 69 other elements considered, with a value of 117.3 kJ/mol taken from the SGTE database \cite{dinsdale_sgte_1991} that is close to that of the most recent thermodynamic assessment of carbon \cite{he_third_2022}, in addition to having the highest first ionization energy. Liquids containing other refractory elements, such as Re, Ir, W and Os, form another cluster (blue) and are expected to show similarities. Liquids between two elements of groups 10 to 16 (yellow) or 1 to 9 (orange) form two distinct clusters, and those between elements of two very different groups are divided into the remaining two clusters (red and green).

It becomes clearer that the most uncertain binary systems belong to particular clusters. In liquids between two elements of similar group (yellow and orange clusters), the uncertainty of the model is relatively low, because 19\% of the compositions are covered in the dataset, and the enthalpy of mixing tend to be of low amplitude \cite{deffrennes_data-driven_2024}. In liquids between two elements of very different groups (red and green clusters), despite also having 18\% of the possible compositions covered in the dataset, the variance is relatively high, because they tend to form complex liquids with strong short-range order and very exothermic and asymmetric ‘V-shaped’ enthalpies of mixing, such as in the Bi-Li system \cite{liu_thermodynamic_2013}. The uncertainty of the model is the highest in liquids containing refractory elements (purple and blue clusters). This can be attributed to a significant lack of data, as only 2\% of the possible composition are covered in the dataset. Indeed, only the C-Ni liquid where the enthalpy of mixing is known in the Ni-rich side \cite{jeon_thermodynamic_2021} is included in our dataset for the carbon cluster, and only 7 binaries are included for the cluster formed by other refractory elements. Therefore, there is a critical need for new data on the liquids formed by refractory elements to capture the trends in their enthalpy of mixing. Although binaries containing C are the most uncertain, we have excluded them from our acquisition plan. This choice is explained later in the section \ref{Discussion}.

\clearpage
\subsection{Data acquisition using ab initio molecular dynamics}\label{AIMD}

In line with our active learning strategy, we focused on obtaining new data in binary liquids between refractory elements. We directed our efforts towards liquids containing W, Os, Ir and Re, given that the uncertainty of the Gaussian process model was particularly high for these elements. Their high melting points make experimental investigations extremely challenging, which explain the scarcity of data in these systems. Therefore, a numerical approach is needed. Ab initio molecular dynamics is the most appropriate simulation technique for accurately calculating the enthalpy of mixing in the liquid phase. It requires large computational resources, but as it will be discussed further, it can provide high quality data in reasonable agreement with experiments.

In order to calculate the enthalpy of mixing in a liquid, three simulations are required: two for the pure elements ($H_A^{\text{pure}}$ and $H_B^{\text{pure}}$) and one for the alloy ($H_{\text{alloys}}$):

\begin{equation}
\Delta H_{\text{mix}} = H_{\text{alloys}} - \left( x_A H_A^{\text{pure}} + x_B H_B^{\text{pure}} \right)
\end{equation}

To circumvent issues related to nucleation, we carefully chose a temperature just above the melting point of the more refractory element of the two, thereby keeping our samples in a liquid phase. Details of our calculations are given in section \ref{subsecAIMD}. Table \ref{tab:mixingenthalpies} shows all the calculated data obtained for equimolar binary liquids. 

\clearpage

\begin{table}
    \centering
    \begin{tabular}{|cc|cc|cc|cc|}
        \hline
        \multicolumn{2}{|c|}{\textbf{Ir alloys}} & \multicolumn{2}{c|}{\textbf{Os alloys}}\\ 
        \hline
        \textbf{System} & \textbf{$\Delta H_{\text{mix}}$ (kJ/mol)} & \textbf{System} & \textbf{$\Delta H_{\text{mix}}$ (kJ/mol)} \\  
        \hline
        Ir-As & -18.89 & Os-As & N/A \\ 
        Ir-Al & -70.40 & Os-Al & -30.31 \\ 
        Ir-Ca & -42.20 & Os-Ca & X \\ 
        Ir-Mg & -33.99 & Os-Mg & X \\ 
        Ir-Mo & -28.61 & Os-Mo & -14.64 \\ 
        Ir-Sc & -86.65 & Os-Sc & -42.71 \\ 
        Ir-Si & -64.06 & Os-Si & -34.65 \\ 
        Ir-Ti & -79.87 & Os-Ti & -46.22 \\ 
        Ir-V & -48.92 & Os-V & -29.72 \\ 
        Ir-Zr & -84.22 & Os-Zr & -42.73 \\ 

        \hline
    \end{tabular}
    \begin{tabular}{|cc|cc|cc|cc|}
        \hline
        \multicolumn{2}{|c|}{\textbf{Re alloys}} & \multicolumn{2}{c|}{\textbf{W alloys}}\\ 
        \hline
        \textbf{System} & \textbf{$\Delta H_{\text{mix}}$ (kJ/mol)} & \textbf{System} & \textbf{$\Delta H_{\text{mix}}$ (kJ/mol)} \\  \hline
        Re-As & N/A & W-As & N/A \\ 
        Re-Al & -8.88 & W-Al & -3.00 \\ 
        Re-Ca & X & W-Ca & X \\ 
        Re-Mg & X & W-Mg & X \\ 
        Re-Mo & X & W-Mo & 1.88 \\ 
        Re-Sc & -12.77 & W-Sc & X \\ 
        Re-Si & -15.13 & W-Si & -15.43 \\ 
        Re-Ti & -18.55 & W-Ti & -3.88 \\ 
        Re-V & -16.64 & W-V & -0.82 \\ 
        Re-Zr & -11.10 & W-Zr & 1.71 \\ 
        \hline
    \end{tabular}
    \caption{Mixing enthalpies obtained by AIMD in equimolar liquids containing Ir, Os, Re and W. Crosses represent binaries for which we observed phase separation. N/A represent system for which calculations were not made.}
    \label{tab:mixingenthalpies}
\end{table}

\clearpage

\subsection{Improved prediction of the enthalpy of mixing in binary liquids}\label{prediction}


We have improved the dataset from Ref. \cite{deffrennes_data-driven_2024} that now includes data on 433 binary liquids, 29 of which are obtained using active learning and AIMD in liquids of refractory elements. On this basis, we update our machine learning model to predict the enthalpy of mixing in binary liquids. We use the same methods as in Ref. \cite{deffrennes_data-driven_2024}: the enthalpy of mixing is obtained indirectly by predicting the error of Miedema’s model \cite{miedema_cohesion_1980} using a LightGBM algorithm. This algorithm is trained on features derived from the properties of the pure element, which are selected by recursive feature selection. 


Model performance is evaluated across the whole dataset using nested cross-validation, in binary systems from which all data was withheld during training and tuning. A root mean squared error (RMSE) of 4.9 kJ/mol is obtained, compared to 7.3 kJ/mol with Miedema’s model. These metrics are not directly comparable to those in Ref. \cite{deffrennes_data-driven_2024} because they are evaluated on different datasets.

In Miedema’s model \cite{miedema_cohesion_1980}, binary liquids are classified into 3 groups: alloys of two transition metals, two non-transition metals, or one transition metal and one non-transition metal, with Zn, Cd and Hg treated as non-transition metals. It is interesting to see how our model performs for each of the 3 groups. The transition metal/transition metal group gives the best performance, with an RMSE of 4.2 kJ/mol, while a higher value of 5.2 kJ/mol is obtained in both the other groups. A map of the groups is given in Supplementary Note 3 to aid understanding. 

\begin{figure}[H]
\centering
\includegraphics[width=0.8\textwidth]{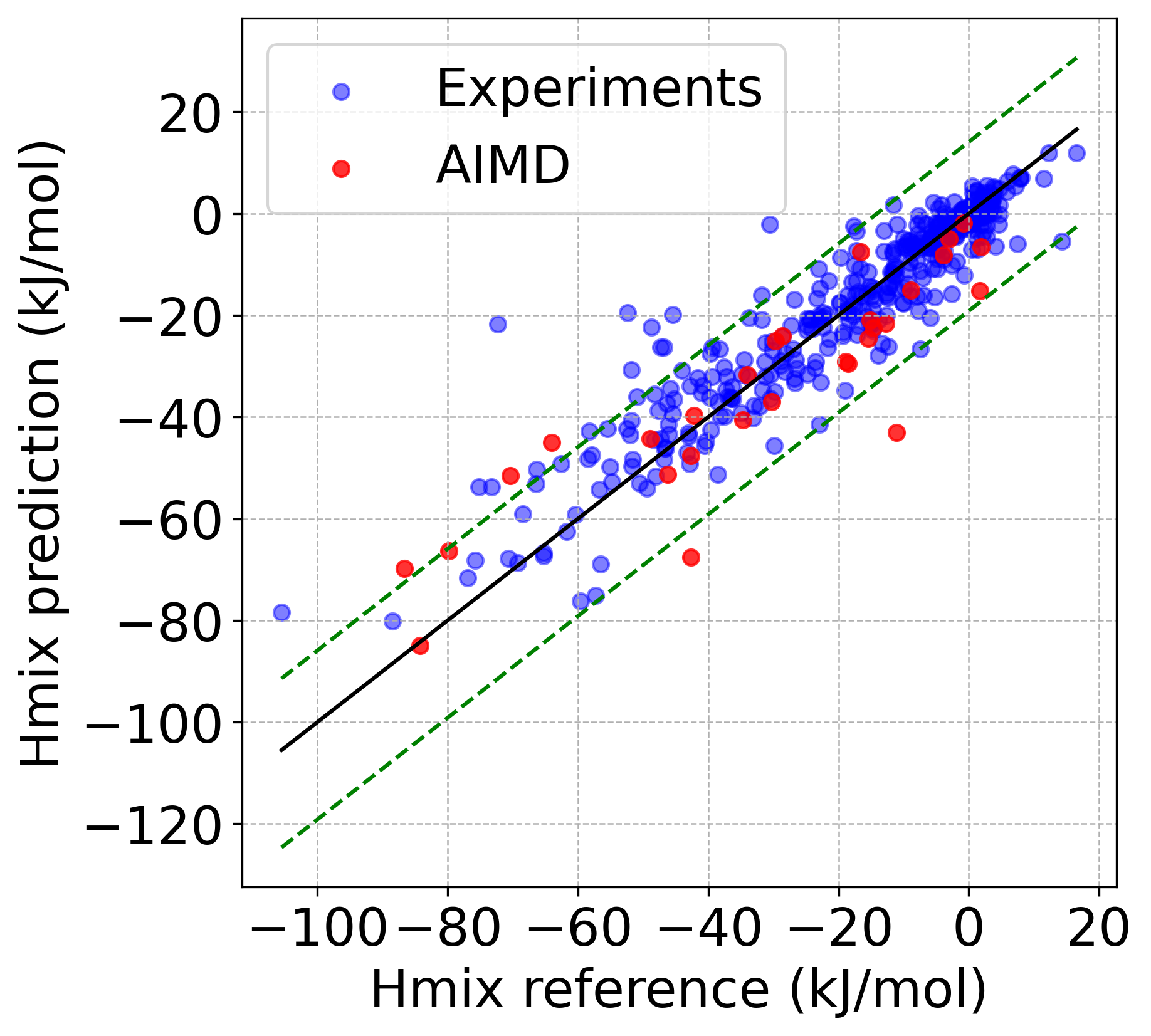}
\caption{Test performance of our LightGBM machine learning model against enthalpy of mixing data obtained by AIMD (red) or dervied from experiments (blue). For clarity, only the data at which the enthalpy of mixing reaches its extremum value is shown for each of the 433 binary system included in the dataset. The black solid line represents a perfect agreement between predictions and observations. 95 percent of the predictions fall within the green dashed lines, which are the 0.025 and 0.975 quantiles.} 
\label{Perf}
\end{figure}

We now focus on the test performance of our model on our 29 AIMD data points from liquids containing Ir, Os, Re and W. This performance is compared to that of a reference model evaluated using the same nested-cross validation procedure, but trained without access to our AIMD data. The global RMSE on AIMD data decreased from 12.8 to 9.5 kJ/mol when AIMD data not used for testing were included in the training set. For a more detailed analysis, the list of RMSE by refractory element is given here:

\begin{itemize}

    \item Os: RMSE decreased from 12.0 (reference model, before AIMD) to 6.7 (final model, after AIMD)

    \item Ir: RMSE slightly improved from 11.4 (before AIMD) to 10.9 (after AIMD)

    \item Re: RMSE dropped from 15.2 (before AIMD) to 12.0 (after AIMD)

    \item W: RMSE significantly decreased from 13.1 (before AIMD) to 7.0 (after AIMD)
\end{itemize}
These results suggest that our strategy has improved prediction accuracy for refractory liquids, particularly for elements such as Os, Re, and W. The improvement is less evident for Ir. AIMD data are highlighted in the parity plot Fig. \ref{Perf}. The results show that performance on these data is consistent to that on the rest of the dataset, except for the Re-Zr, Os-Zr, and W-Zr systems, for which model predictions tend to be more exothermic than AIMD calculations. Conversely, the Ir-Si, Ir-Sc, Ir-Ti, and Ir-Al systems are predicted to be less exothermic than their AIMD-calculated values. These discrepancies suggest potential systematic biases in the model, particularly for Zr-containing refractory systems and Ir-based alloys.

\begin{figure}[H]
\centering
\includegraphics[width=0.8\textwidth]{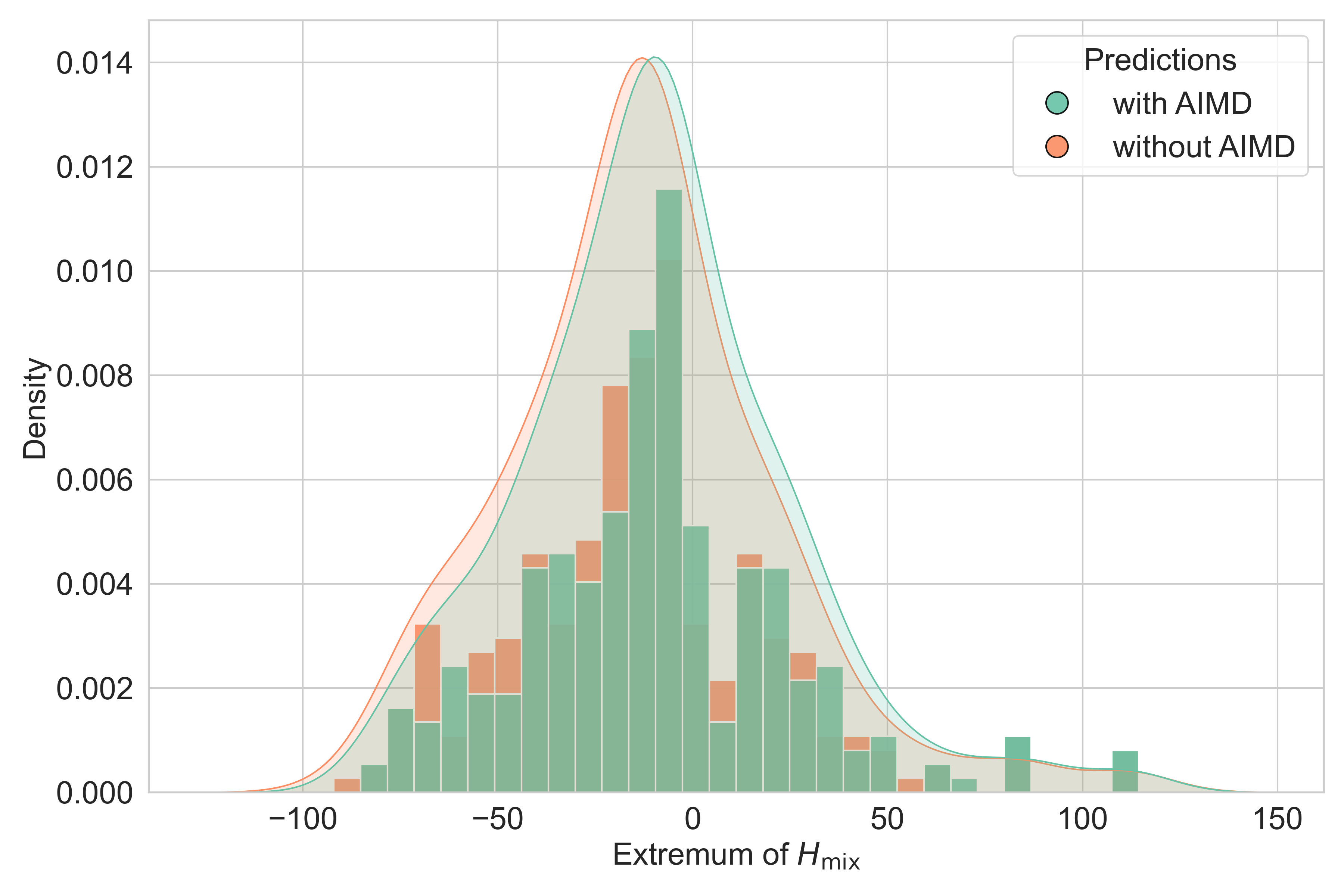}
\caption{Probability density of the extremum value of the enthalpy of mixing in equimolar binary liquid containing Ir, Os, Re and W before (orange) and after (green) the inclusion of our AIMD data. Histogram represents our predictions.}
\label{Figurecomparison}
\end{figure}

The addition of new data has enabled our model to correct predictions that were in general too exothermic in liquids containing refractory elements. 
Fig. \ref{Figurecomparison} illustrates this effect with the probability density of the extremum value of the enthalpy of mixing over Ir, Os, Re, and W equimolar liquids, where we can see a shift of the data toward less negative values, which is in average of +3.7 kJ/mol. This shift seems a little more pronounced for highly exothermic alloys. This correction can explain why our prediction in test conditions for Ir alloys are a little further away than their reference values.

\clearpage
\subsection{Trends in the enthalpy of mixing for refractory elements}\label{refractory}

\begin{figure}[h]
\centering
\includegraphics[width=0.9\textwidth]{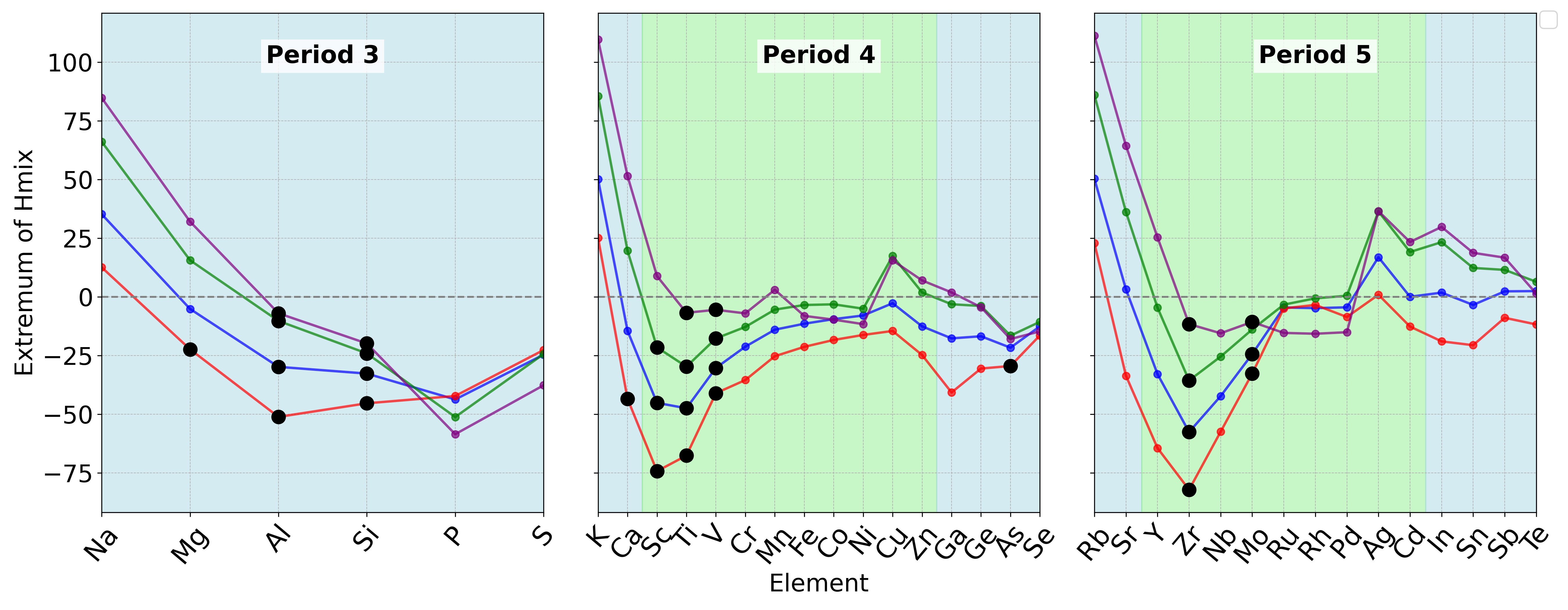}
\caption{Extremum value of enthalpy of mixing in binary liquids between W (purple), Re (green), Os (blue), or Ir (red) as a function of the second element from periods 3, 4, and 5 sorted by atomic number. Black points represent the systems in which AIMD calculations were made. A blue background represents the liquids between a transition metals with a non-transition metals, whereas a green background represents the liquids between two transition metals.}
\label{Trends}
\end{figure}

To understand the general trends predicted by our model for the refractory elements of period 6, Fig.\ref{Trends} show the extremum value of the enthalpy of mixing in binary liquids composed of W, Re, Os, or Ir and a second element from periods 3, 4, and 5. In general, interactions become more attractive (or less repulsive) as the atomic number of the refractory element increases from W to Ir. This is consistent with our AIMD calculations that are highlighted by black circles on the figure. Interactions with alkali metals, and to a lesser extent alkaline earth metals, are strongly repulsive. This is consistent with the observations of large miscibility gaps in these systems. According to Miedema’s theory, this can be explained by their similar electronegativity, but very different electron densities, compared to refractory elements. Interactions are very attractive with early transition metals of group 3 to 5, and the enthalpy of mixing becomes progressively less exothermic as the group number increases across the transition metal series, sometimes even endothermic with noble metals, until a plateau is reached with elements of group 13 and above. This is again consistent with phase diagram observations, such as the formation of compounds and solid solutions in the Ir–Ti system \cite{wang_thermodynamic_2014}, as opposed with the absence of miscibility in alloys between noble metals and W \cite{vijayakumar_calculated_1988}. Fig. \ref{Trends} demonstrate the relevance of our data acquisition strategy, as the new AIMD data were acquired in systems where the enthalpy of mixing reaches its most exothermic values.

\clearpage
\subsection{Heat capacity and column importance in the framework of Miedema theory}

\begin{figure}[H]
\centering
\includegraphics[width=0.9\textwidth]{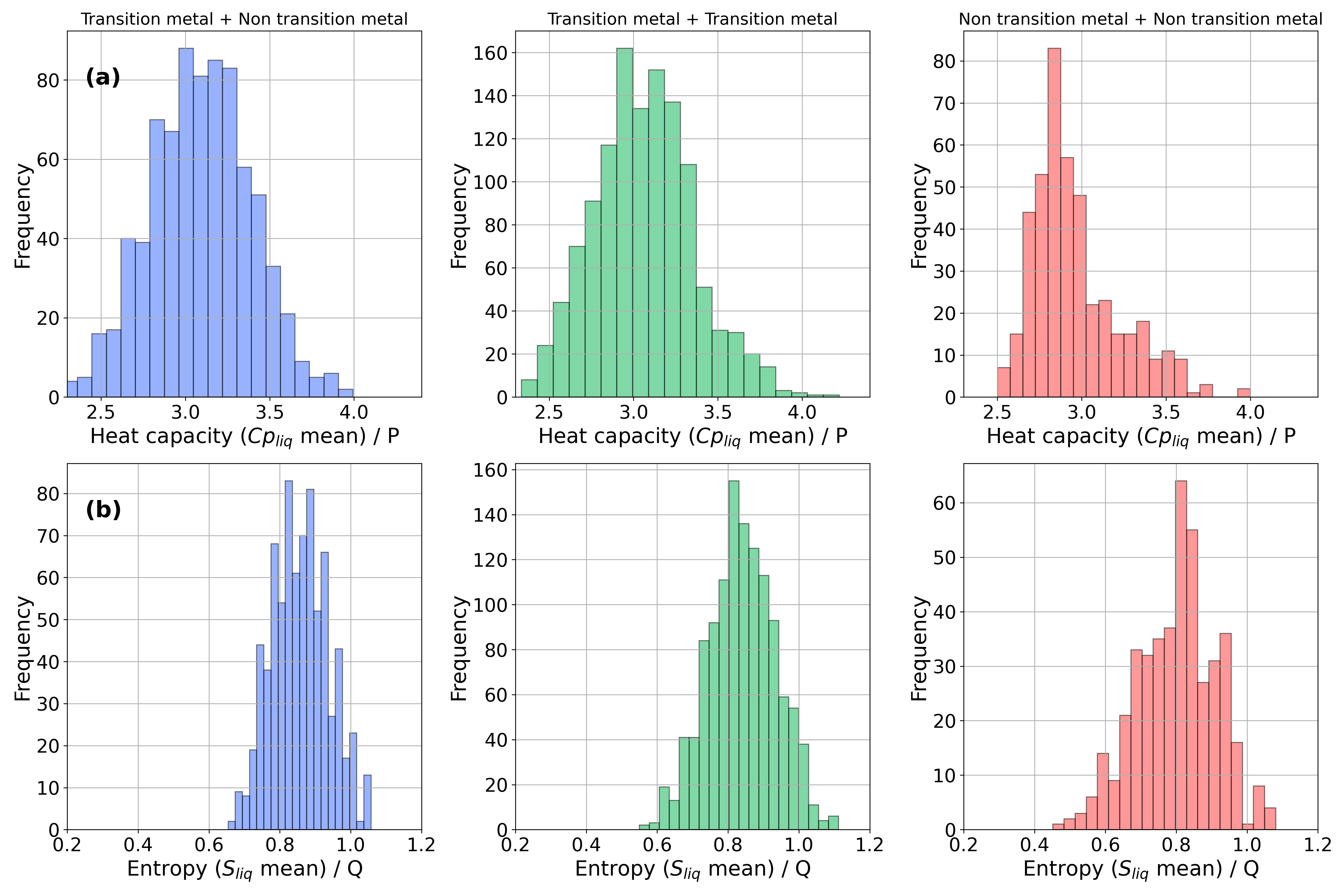}
\caption{Histograms of (a) the mean heat capacity between elements just above their melting point divided by the P parameter of Miedema’s model, and (b) the mean entropy between element just above their melting point divided by the Q parameter of Miedema’s model for (left) liquids between two transition metals, (middle) liquids between a transition metal and a non-transition metal, and (right) liquids between two non-transition metals.}
\label{histosclusterscpliq}
\end{figure}

As discussed in Section 2.3, in Miedema’s model classifies binary liquids into three groups based on whether the elements are transition metals. Predictions are obtained from two empirical parameters, Q and P, with P equals to 14.2 for alloys of two transition metals, 10.7 for alloys of two non-transition metals, and 12.35 otherwise, while Q is defined as $Q = 9.4 \times P$. We found that we can closely reproduce Miedema’s three group using k-means clustering on two of the most important features in our model: the mean heat capacity calculated for both elements in the liquid at the melting point, and their mean entropy calculated in the same manner. A clustering accuracy of 79\% was obtained, indicating that 79\% of the data points were correctly assigned to the corresponding clusters of Miedema, as shown in Suppl. Note 3 (Fig. S5). It appears from Fig. \ref{histosclusterscpliq} that P is roughly equal to three times the mean heat capacity between elements, while Q is roughly equal to 0.8 times the mean entropy. This result suggests that these parameters could be replaced in Miedema’s model by quantities related to the heat capacity and entropy of liquids, which is a potential direction for improvement.

\clearpage
\section{Discussion}\label{Discussion}

We implemented an active learning strategy that enabled us to identify in which systems data collection was most relevant for predicting the enthalpy of mixing in the liquid phase. Binary liquids containing refractory metals proved to be the among systems with the highest prediction uncertainty. For these reasons, we focused on W-, Re-, Ir- and Os-based alloys. Data acquisition for 29 systems was then carried out by AIMD calculations. These simulation results have been added to the a dataset which is based on experimental measurements, raising the question of the consistency between the two data sources.
To address this question, we reviewed the literature for liquids with both AIMD and experimental data available. We only found 5 binary systems (Al-Si \cite{qin_structure_2016}, Al-Ti \cite{xu_mixing_2022}, Ni-Ti \cite{li_structural_2023}, Al-Fe \cite{han_accurate_2013} and Mg-Si \cite{wang_correlation_2021}), as there are only a few studies where AIMD simulations are used specifically to calculate the enthalpy of mixing due to the large computational resources required. We find that the data obtained by AIMD remain relatively close to the experiments, with an average absolute error of 4.6 kJ/mol, disregarding temperature differences. Our AIMD calculations allowed us to refine the predictions of our machine learning model for refractory alloys, which were in general too exothermic. Close attention to the interpretation of our model using clustering, trend, and feature analyses further support the validity of our approach. 

The enthalpy of mixing in the liquid is known to decrease in absolute value when the temperature increase \cite{kaptay_tendency_2012}. A limitation of our model is that it is trained on data obtained a different temperature, but does not account for this dependence. Most of our data are based on experiments that are usually carried out close to the liquidus in the system, and below 2000K. Our AIMD data were obtained at particularly high temperatures, as calculations were performed above the melting point of the most refractory element so that it can be used as a reference in the liquid state. We initially conducted calculations on C-based liquids where prediction uncertainty is the highest. However, simulations had to be performed above the high melting temperature of C to avoid nucleation, and the enthalpy of mixing values we obtained turned out to be zero almost systematically, which may not be the case at lower temperatures. An important, but challenging perspective to this work is to take into account the temperature at which experiments or simulations were made to refine the predictions. At present, our knowledge of the temperature dependence of the enthalpy of mixing remains limited, as data are scarce and only available at relatively low temperatures \cite{deffrennes_data-driven_2024}.

Our model can predict an asymmetry in the enthalpy of mixing with respect to composition, unlike Miedema's model, but with an accuracy that remains limited \cite{deffrennes_data-driven_2024}. This asymmetry generally originates from short-range local ordering. Several studies have demonstrated the ability of AIMD simulations to accurately capture such local interactions. However, we could not improve the model on this aspect, which appears challenging due to the large number of simulations required to account for these effects. A possible alternative would be to identify a relatively robust and widely available feature across binary systems that could capture short-range interaction effects.

\pagebreak
\section{Methods}\label{Methods}


\subsection{Data collection}\label{datacollection}

In our previous work \cite{deffrennes_dataset_2024}, we collected data on the enthalpy of mixing in 375 binary liquids from Calphad assessments at compositions where the models are supported by experiments. In this study, we expanded this dataset with data supported by direct calorimetric measurements found in 29 new binary liquids \cite{ivanov_enthalpies_2018, fels_calorimetric_2019, onderka_thermodynamic_2013, wang_experimental_2017, gasior_thermodynamic_2018, guo_thermodynamic_nodate, luef_enthalpies_2004, elmahfoudi_enthalpy_2012, jendrzejczyk-handzlik_thermodynamic_2012, shevchenko_thermodynamic_2015, kumar_thermodynamic_2021, shevchenko_thermodynamic_2016, moser_cdga_1988, kang_critical_2007, kim_critical_2014, morachevskii_liquid_2014, ivanov_enthalpies_2017, kim_critical_2012, sudavtsova_thermodynamic_2017, usenko_enthalpies_2020}.

\subsubsection{Machine learning}\label{ML}

For clustering, we used the kmeans algorithm implemented in scikit-learn \cite{pedregosa_scikit-learn_2011}. In order to obtain the optimal number of groups K, we used the elbow method with the silhouette method as detailed in Supplementary Note C.
For active learning, we used the random forest algorithm implemented in scikit learn \cite{pedregosa_scikit-learn_2011}, and the Gaussian process algorithm implemented in Physbo \cite{motoyama_bayesian_2022}. Feature selection was made using recursive feature elimination as detailed in Supplementary Note A.
For the final prediction model, we used the LightGBM algorithm \cite{ke_lightgbm_2017}. The procedure is the as described in our previous work \cite{deffrennes_data-driven_2024} and is only briefly summarized here. It is based on a nested cross-validation (CV) in which data on a binary system form a group that cannot be split between the training, validation and test sets. In the inner loop, a 11-fold CV is performed where feature are selected from a recursive feature elimination, and hyperparameters are tuned using Optuna \cite{akiba_optuna_2019}. In the outer loop, a 12-fold CV is performed to evaluate model performance on the whole dataset.

\subsection{Ab Initio Molecular Dynamics}\label{subsecAIMD}

Ab Initio Molecular Dynamics simulations were performed using the code Vienna Ab Initio Simulation Package (VASP) \cite{kresse_ab_1993} by considering constant numbers of atoms N, volume V, and temperature T (NVT), namely the canonical ensemble \cite{nose, hoover_canonical_1985}. The dynamics were carried out by numerically solving Newton's equations of motion using the velocity form of the Verlet algorithm, with a timestep of 3 fs. The electronic system was treated with electron-ion interactions represented by the projector augmented wave (PAW) potentials \cite{kresse_ultrasoft_1999, blochl_projector_1994}. Correlation effects were taken into account with the Generalized Gradient Approximation (GGA) \cite{wang_correlation_1991} in the Perdew, Burke, and Ernzerhof (PBE) \cite{perdew_self-interaction_1981} formulation. The plane-wave expansion utilized a cutoff energy of 400 eV. This choice was made so as to be sufficiently broad to maintain homogeneity throughout the simulations. Due to the large supercells and the liquid state, Brillouin zone sampling was restricted to the $\Gamma$ point.

The total number of atoms is 100 for all simulations. The temperature at which simulations are run is a little above the melting temperature of the most refractory element, to avoid any solidification complications. For this reason, the temperatures we have considered for Ir, Os, Re and W alloys are 2750, 3300, 3500 and 3700 K, respectively. As the ensemble considered was the NVT, we adjusted the volume so that the pressure oscillated around a mean value of between -2 and 2 kbar, which enables us to neglect the pV term. Once the volume has been adjusted, the simulation time is at least 10 ps to extract the thermodynamic quantities. In cases where the system is not sufficiently converged, a longer time is taken to ensure that it is suitably relaxed. Particular care is taken to ensure that simulations are in the liquid state, by analyzing structural indicators and the partial pair correlation function. This procedure has been applied to all systems in the Table \ref{tab:mixingenthalpies} to obtain the resulting enthalpy of mixing in the liquid.

\pagebreak
\section*{Declarations}

\section*{Data Availability}

Our dataset is available in an open-access data repository \cite{deffrennes_dataset_2024}. This repository also includes predictions of the enthalpy of mixing from our machine learning model for all the 2415 binary liquids shown in Fig. \ref{fcov}. These predictions are given as Redlich-Kister polynomials following the same fitting procedure as Ref. \cite{deffrennes_data-driven_2024} so that they can be readily used in Calphad assessments or extrapolation models for multicomponent liquids.

\section*{Acknowledgements}

This work has been partially supported by MIAI@Grenoble Alpes, (ANR-19-P3IA-0003), and by a France 2030 government grant managed by the Agence Nationale de la Recherche with the reference ANR-23-IACL-0006. This work was partially supported by JSPS KAKENHI Grant Number 25K01492. Part of computations presented in this paper were performed using the GRICAD infrastructure (https://gricad.univ-grenoble-alpes.fr), which is supported by Grenoble research communities. We acknowledge the CINES for computational resources under project number AD010914852 and AD010914852R1. We thank François Bottin and Romuald Béjaud (LMCE, CEA-Univ. Paris-Saclay) as well as Noel Jakse (SIMaP, Univ. Grenoble Alpes-CNRS-Grenoble INP) for valuable discussions regarding AIMD simulations.

\noindent

\bigskip
\begin{flushleft}%


\end{flushleft}%

\end{document}


\title{Active Learning for Predicting the Enthalpy of Mixing in Binary Liquids Based on Ab Initio Molecular Dynamics}


\author[1,a]{Quentin Bizot\thanks{quentin.bizot@rub.de}}
\author[2,3]{Ryo Tamura\thanks{tamura.Ryo@nims.go.jp}}
\author[1]{Guillaume Deffrennes\thanks{guillaume.deffrennes@cnrs.fr}}

\affil[1]{Univ. Grenoble Alpes, CNRS, Grenoble INP, SIMaP, F-38000, Grenoble, France}
\affil[2]{Center for Basic Research on Materials, National Institute for Materials Science, 1-1 Namiki, Tsukuba, Ibaraki, 305-0044, Japan}
\affil[3]{Graduate School of Frontier Sciences, The University of Tokyo, 5-1-5 Kashiwa-no-ha, Kashiwa, Chiba, 277-8561, Japan}
\affil[a]{\textit{Present address of Q. Bizot:} Interdisciplinary Centre for Advanced Materials Simulations (ICAMS), Ruhr Universität Bochum, Bochum, Germany}

\maketitle

\pagebreak


\section{Feature selection}\label{secA1}

\subsection{Feature selection for active learning}\label{FeatureAL}
Feature selection was performed using the Recursive Feature Elimination (RFE) method, which is applicable to random forests. Before each elimination, we trained a LightGBM model 25 times and removed the less important feature in average. To mitigate statistical variability, we repeated the feature evaluation 60 times with different splits of our dataset. The root mean square error (RMSE) as a function of the number of features is shown in the Fig. \ref{featureselection}. For active learning, we selected the set of ten features that appeared most frequently across the repeated trials and that are presented in the results section of the article. We chose 10 features as a compromise between accuracy, interpretability, and computational time.

\subsection{Selected feature of our final LightGBM model}\label{FeaturelightGBM}


\begin{figure}[h]
\centering
\includegraphics[width=0.9\textwidth]{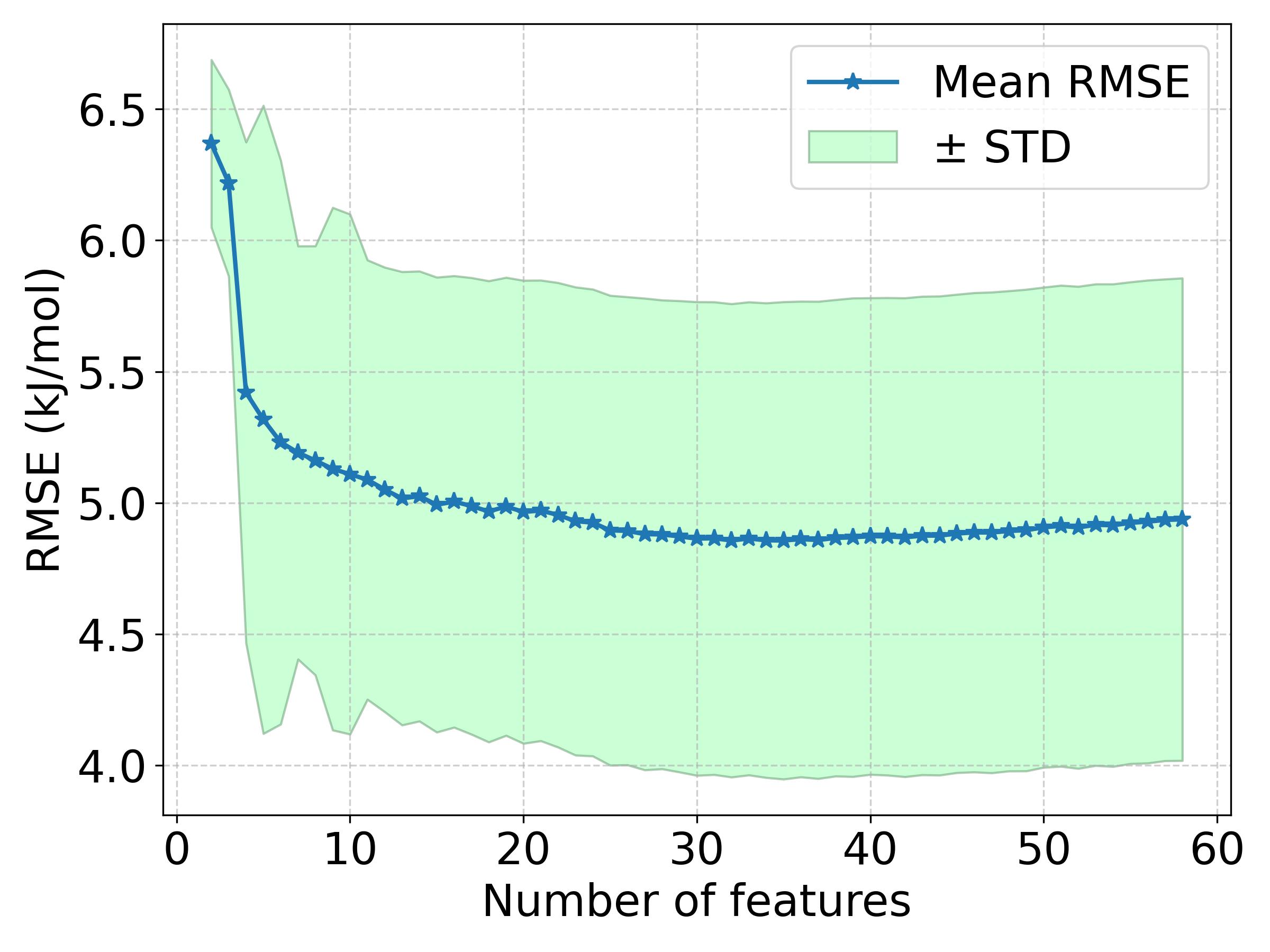}
\caption{Feature selection using Recursive Feature Elimination (RFE). Mean of the Root Mean Square Error (RMSE) is plotted in blue color together with the standard deviation in green color in function of the number of features.}
\label{featureselection}
\end{figure}

Features used to train the model are listed in the Table \ref{tablefeature}. These 30 features are chosen as the most important according the Recursive Feature Elimination (RFE) shown in Fig. \ref{featureselection}.

\begin{table}[h!]
\centering
\begin{tabular}{|c|l|}
\hline
\textbf{N°} & \textbf{Features} \\
\hline
1  & ThermalConductivity avg dev \\
2  & Electronegativity avg dev \\
3  & Sfus mean \\
4  & Sliq avg dev \\
5  & NValence avg dev \\
6  & NUnfilled mean \\
7  & MendeleevNumber avg dev \\
8  & MeltT avg dev \\
9  & Hfus avg dev \\
10 & Sfus avg dev \\
11 & FirstIonizationEnergy avg dev \\
12 & Electronegativity mean \\
13 & HeatVaporization mean \\
14 & Polarizability mean \\
15 & Cpfus mean \\
16 & Column avg dev \\
17 & Column mean \\
18 & Cpliq mean \\
19 & Cpliq avg dev \\
20 & Cpfus avg dev \\
21 & Density avg dev \\
22 & Density mean \\
23 & Ssol mean \\
24 & MendeleevNumber mean \\
25 & Cpsol avg dev \\
26 & Hsol mean \\
27 & BoilingT avg dev \\
28 & MolarVolume avg dev \\
29 & Hliq avg dev \\
30 & HeatVaporization avg dev \\
\hline
\end{tabular}
\caption{List of features used to train the final LightGBM model}
\label{tablefeature}
\end{table}

\clearpage
\section{Active Learning benchmark}\label{featselec}

\begin{figure}[h]
\centering
\includegraphics[width=0.9\textwidth]{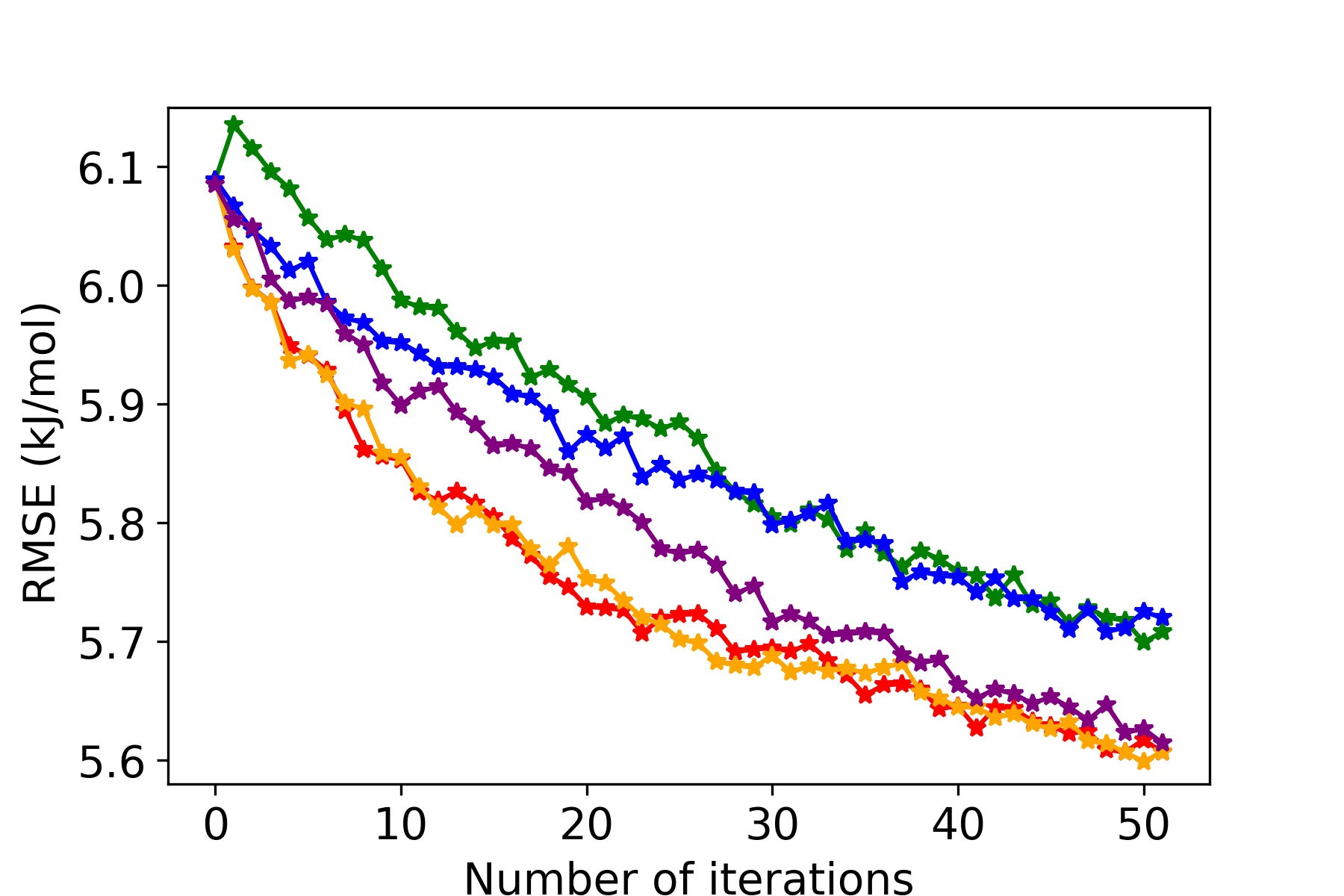}
\caption{RMSE (kJ/mol) in function of the number of iterations. Each iteration represents the addition of 4 points (i.e. a mixing enthalpy value to a composition) depending on the algorithm used. Each iteration represents the addition of 4 points (i.e. a mixing enthalpy value to a composition) depending on the algorithm used. The algorithms used are detailed in the text by the different items Approach 1, Approach 2, Approach 3, Approach 4, and Approach 5, which are represented on the curve by the colors green, blue, violet, red and yellow, respectively.}
\label{FigureActiveLearning}
\end{figure}

Different active learning approaches are tested to find the most efficient one. The dataset is divided into a training containing data on 100 randomly selected binary systems (about 25 percent of our data), a test set containing data on another 100 binaries, and a pool set containing data on the remaining 204 binaries. Data are iteratively selected from the pool set using a specified active learning approach and added to the training set to update a LightGBM model, whose performance is evaluated on the test set. To ensure statistical robustness, each active learning approach is evaluated on the same 100 predefined random splits of the data into training, test, and pool sets.


\begin{itemize}
    \item Approach 1: 4 binary systems are randomly selected from the pool at each iteration. We add to the training set the data for the equimolar liquid, or, if unavailable, the data from nearest composition. Once a binary has been selected, it is removed from the pool to prevent future selection. This approach, which is similar to random sampling, is considered the benchmark against which other algorithms are compared. It is shown in green in Fig. \ref{FigureActiveLearning}.
    \item Approach 2: It is based on the regression tree–based active learning (RT-AL) approach proposed in the literature that uses on regression trees capable of capturing the diversity in the data in order to obtain a better representativeness in the selection. 4 points are selected at each iteration, and unlike the first approach, multiple points from the same binary can be selected across iterations. This method, shown in blue in Fig. \ref{FigureActiveLearning}, improves RMSE on the first iterations, but finally returns to the RMSE values obtained with random sampling on the last iterations. This is because, in aiming for diversity in the data, the selection process tends to favor data at dilute compositions near the pure elements, which provide limited information as the enthalpy of mixing is then close to zero.
    \item Approach 3: Since the second algorithm failed to identify relevant compositions, we relied solely on RT-AL to select 4 binary systems at each iterations, from which we selected data for the equimolar liquid (or the closest composition available) and subsequently removed them from the pool set. The purple curve in Fig. \ref{FigureActiveLearning} shows that this strategy is more interesting, as it results in a more significant reduction in RMSE than Approach 2.
    \item Approach 4: Gaussian Process is used to select the most uncertain points based on covariance. As in Approach 2, four points are selected at each iteration, and the same binary system can be sampled several times. The curve associated with this algorithm is the yellow one (Fig. \ref{FigureActiveLearning}), where we observe a sharp decrease in RMSE at the beginning. However, in the final iterations, the RMSE values converge with those of Approach 3. This may come from the fact that the fact that the pool set is limited as it includes only 50 percent of our dataset, which only covers 17 percent of all binary alloys.
    \item Approach 5: We tested whether the acquisition procedure based on Gaussian Process would improve by discarding binaries once a point was selected as done in Approach 3. As shown by the red curve (Fig. \ref{FigureActiveLearning}), this does not improve RMSE values, as the curves are nearly identical.
    
\end{itemize}

For all the reasons mentioned earlier, the approaches that reduce RMSE the most with the fewest iterations are Approaches 4 and 5, both based on the Gaussian Process. The selected algorithm for this study is the Gaussian Process of Approach 4.

\pagebreak

\section{Kmeans clustering}\label{Kmeans}

\subsection{Inertia and sihlouette score}\label{Inertiaandsihlouettescore}

\begin{figure}[h]
\centering
\includegraphics[width=0.9\textwidth]{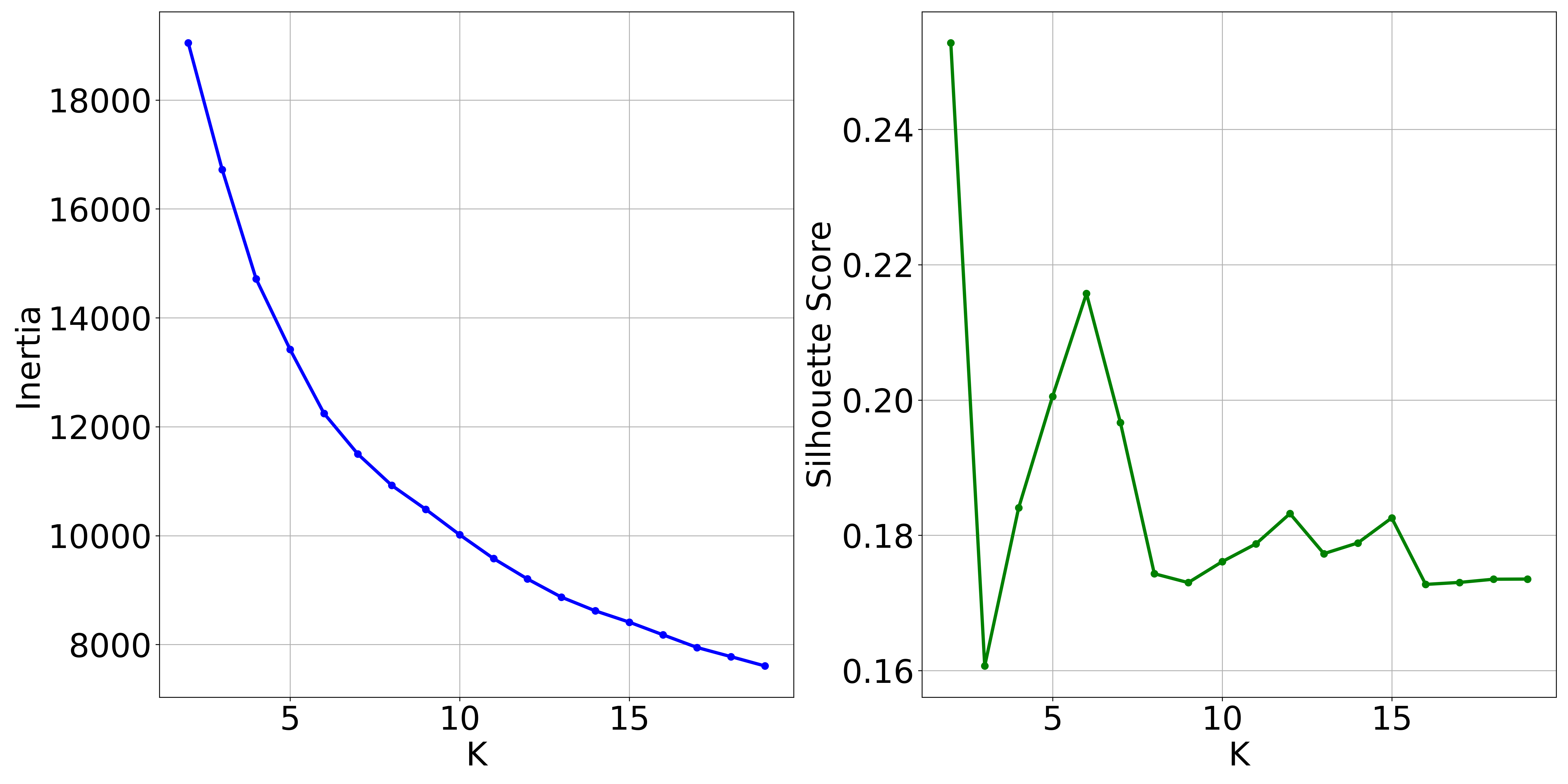}
\caption{Kmeans clustering with the 10 most important features. Left, inertia in function of the number of K clusters. Right, silhouette score in function of K clusters.}
\label{silandiner10features}
\end{figure}

This section is dedicated to the different Kmeans we used in this work and to give additional information about the silhouette and inertia. Fig. \ref{silandiner10features} shows the inertia and the silhouette score considering the 10 most important features used in our active learning strategy. The silhouette score indicates that the optimal number of K clusters to be 2. However, in the framework of our active learning strategy, it seems to give better correlation with Gaussian process with the second peaks, e.g. K=6.

\subsection{Entropy and heat capacity clustering}\label{entropyandheatcapclustering}

\begin{figure}[h]
\centering
\includegraphics[width=0.9\textwidth]{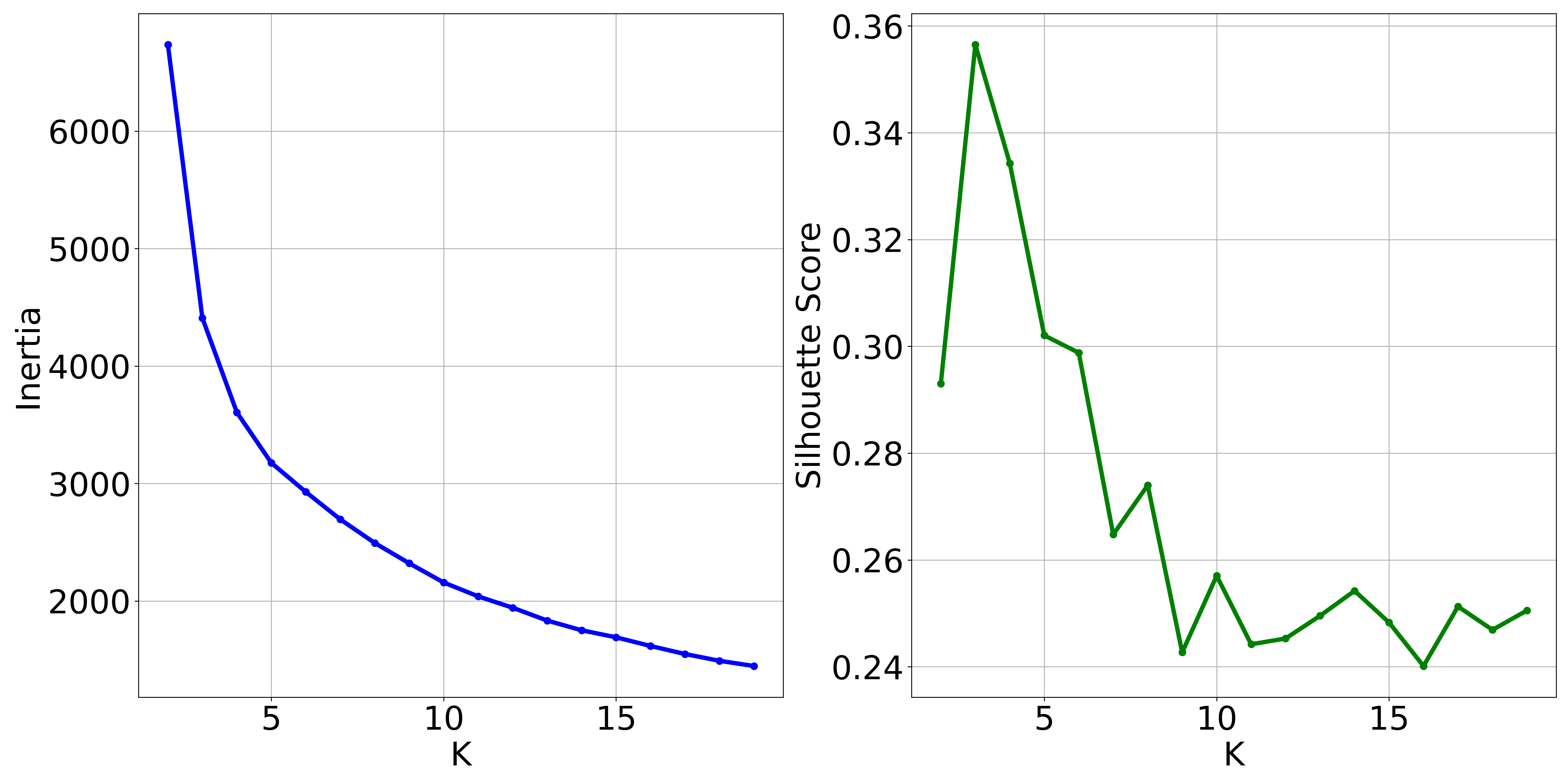}
\caption{Kmeans clustering with the average deviation and mean of the heat capacity and entropy in the liquid phase. Left, inertia in function of the number of K clusters. Right, silhouette score in function of K clusters.}
\label{silandinerCpliqSliq}
\end{figure}

Fig. \ref{silandinerCpliqSliq} shows the inertia and the silhouette score considering the average deviation and the mean of the heat capacity and the entropy in the liquid phase. In contrast to Kmeans with 10 features, we can see a very well-defined peak for K=3 indicating that this number of clusters is optimal if we consider heat capacity and entropy.

\begin{figure}[h]
\centering
\includegraphics[width=0.9\textwidth]{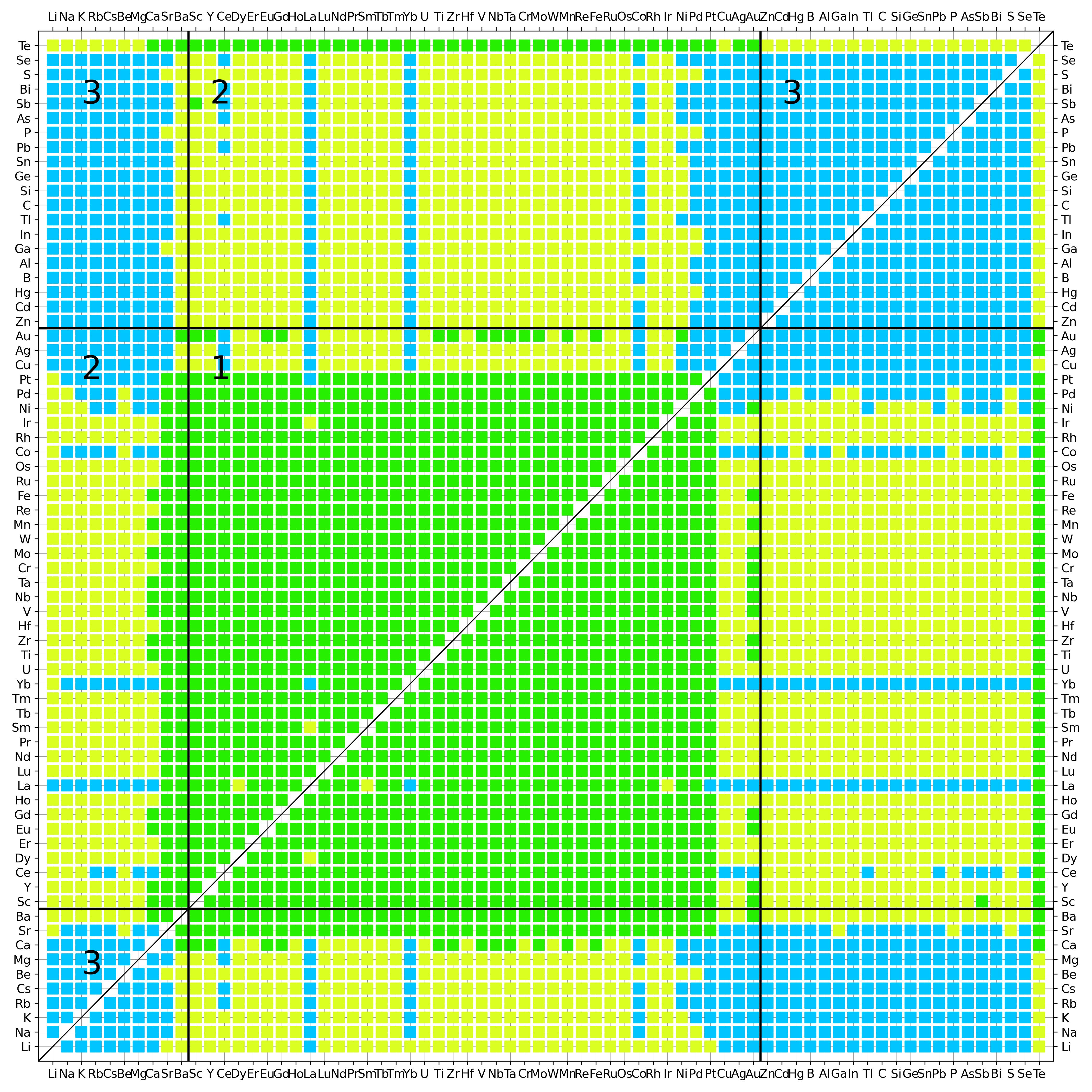}
\caption{Kmeans clustering with the average deviation and the mean of the heat capacity (Cp) and the entropy (Sliq) in the liquid. Black solid lines in the map represent limits between the different group of Miedema with 1 the metals-metals, 2 metals-non metals and 3 non metals- non metals.}
\label{Clusters}
\end{figure}

Fig. \ref{Clusters} shows the resulting map of the three clusters considering the mean heat capacity and entropy. There is a great similarity between the Miedema groups and our kmeans, where blue is close to the transition metal- transition metal group (1), yellow to the transition metal-non-transition metal group (2) and green to the non-transition metal-non-transition metal group (3).

